\documentclass[11pt]{article}

\usepackage[T1]{fontenc}
\usepackage[utf8]{inputenc}
\usepackage{lmodern}
\usepackage[margin=1in]{geometry}
\usepackage{microtype}

\usepackage{graphicx}
\usepackage{booktabs}
\usepackage{subcaption}
\usepackage{amsmath}
\usepackage{amssymb}
\usepackage[table]{xcolor}
\usepackage{pdfpages}

\usepackage[numbers,sort&compress]{natbib}
\usepackage[hidelinks]{hyperref}
\usepackage{url}

\usepackage{setspace}
\newcommand{\Endparasplit}{}
\newcommand{\dropcap}[1]{#1}
\newcommand{\dataavail}[1]{%
  \section*{Data Availability}
  #1
}

\begin{document}

\title{\textbf{Political Sorting Can Drive AI Models Apart Through User Feedback}}

\author{
Petter T\"ornberg$^{1,*}$,
Michael Heseltine$^{2}$,
Nicolò Pagan$^{3}$,\\
Christopher Bail$^{4}$,
Michelle Schimmel$^{1}$,
Christopher Barrie$^{2,5}$
\\[0.8em]
\small $^{1}$Institute for Logic, Language and Computation, University of Amsterdam, Netherlands\\
\small $^{2}$Department of Sociology, University of Oxford, Oxford, UK\\
\small $^{3}$Department of Informatics, University of Zurich, Zurich, Switzerland\\
\small $^{4}$Department of Sociology, Duke University, Durham, NC 27516, USA\\
\small $^{5}$Department of Sociology, New York University, NY 10012, USA\\[0.6em]
\small $^{*}$Corresponding author: \href{mailto:p.tornberg@uva.nl}{p.tornberg@uva.nl}
}

\date{\today}

\maketitle

\begin{abstract} 
\noindent Large language models are rapidly becoming an important source of political information. This raises a fundamental question: will AI systems support a shared basis for political knowledge, or lead different political groups to rely on increasingly different models? Political sorting can drive model fragmentation if three conditions hold: politically different users select into different models, learning from user feedback pushes those models apart politically, and the resulting differences shape subsequent model choices. We call this self-reinforcing process the \textit{centrifugal alignment spiral}. We study its components in three steps. First, we draw on a human experiment showing that political identity predicts model choice. Second, we fine-tune language models on synthetic feedback reflecting predominantly Democratic or Republican preferences. Across five independent runs per model family, paired models diverged on 12--41\% of unseen survey questions with large partisan gaps, and in every run the differences moved in the expected political direction; for some models, differentiation extended even to issue areas excluded from training. Pooling feedback across groups instead suppressed divergence. Third, an empirically anchored agent-based model shows what follows when political sorting and model adaptation operate together: models attract politically distinct audiences, learn from them, and diverge further. User feedback can therefore turn political sorting among AI users into durable differences between the models on which they rely for political information.
\end{abstract}

\noindent\textbf{Keywords:}
artificial intelligence; political identity; large language models; political sorting; human feedback

\vspace{1em}




\vspace{1.5em}
\newpage
\section*{Introduction}
\noindent\dropcap{T}he fragmentation of the political information environment has been one of the defining concerns of the social-media era, encompassing debates over audience sorting, echo chambers, misinformation, polarization, and the erosion of shared reality \citep{Sunstein2001,Flaxman2016,Cinelli2021,Tornberg2022,GonzalezBailon2023}. Large language models (LLMs) may appear to reverse this logic: a small number of general-purpose systems can synthesize information across sources and perspectives, and recent work suggests that they can identify common ground and correct deeply held false beliefs \citep{Tessler2024,Costello2024,argyle2023leveraging}. In principle, LLMs could support a more integrated information environment and a shared basis for political knowledge. Yet their adaptivity also introduces a novel route to fragmentation: if politically different users choose different models, and each model is refined on feedback from its own audience, then political sorting can drive those systems apart. As model differences grow, they may in turn reinforce subsequent sorting. We call this self-reinforcing process the \emph{centrifugal alignment spiral}.

The mechanism joins two processes that have rarely been studied together. Political communication research shows that people often prefer congenial sources and evaluate information through partisan and ideological lenses \citep{Campbell1960,Converse1964,Taber2006,Mutz2006,Garrett2009,Stroud2011,iyengar2009red,heseltineajps}, creating the potential for political groups to sort across sources in fragmented media environments \citep{Sunstein2001,Flaxman2016,Kubin2021,Suiter2021,Shehata2022}. Machine-learning research has long shown that deployed systems can reshape the data on which they subsequently learn \citep{Chaney2018,Jiang2019,Pagan2023FeedbackLoops,Perdomo2020Performative}. This dynamic is especially relevant for modern LLMs, which are repeatedly refined using human feedback and other forms of preference-based alignment \citep{Christiano2017,Ziegler2019,Ouyang2022,Bai2022Helpful,Rafailov2023}. As alignment becomes increasingly tailored to particular users and populations, a growing literature has begun to examine group-specific, personalized, and in-situ approaches \citep{Zhao2024GPO,Wang2025UserFeedbackAlignment,Shi2026WildFeedback}. Our contribution is to show that \Endparasplit \noindent these two processes can couple: political sorting may determine which audiences provide feedback to competing systems, making audience composition a cause of model behavior. Unlike fragmentation around common content or public networks, this process can emerge through separate, often private interactions with adaptive systems.

\begin{figure}[t!]
\centering
\includegraphics[
width=0.7\columnwidth,
trim={4.0cm 6.0cm 5.2cm 2.8cm},
clip
]{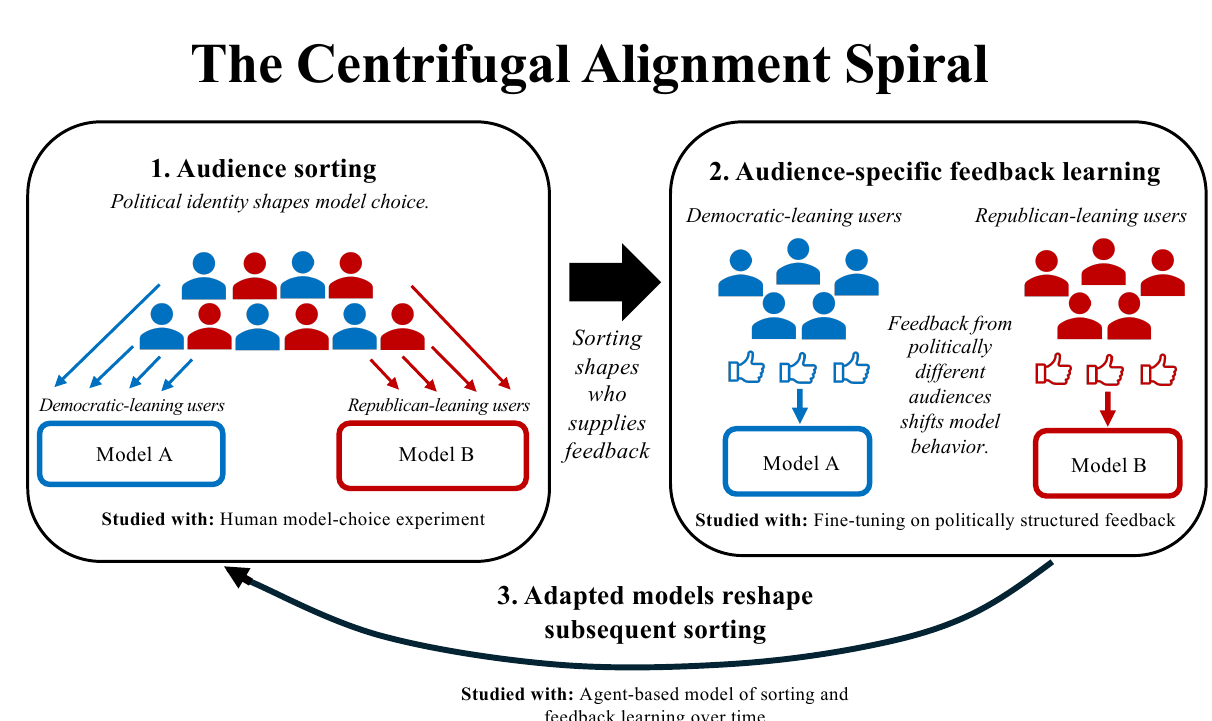}
\caption{\textbf{The hypothesized centrifugal alignment spiral and study
design.} Political identity can shape model choice, determining which
audiences provide feedback to competing systems. Feedback from politically
different audiences may then differentiate model behavior, potentially
reshaping subsequent model choices. We study audience sorting using a human
model-choice experiment, feedback learning using fine-tuning on politically
structured feedback, and the dynamic coupling of these processes using an
empirically anchored agent-based model.}
\label{fig}
\end{figure}

Whether the centrifugal alignment spiral becomes self-reinforcing is an open empirical question. Three conditions must hold: (1) political identity must shape model choice; (2) audience-specific feedback must produce persistent model differences that extend beyond the precise examples used for updating; and (3) those differences must become salient enough to affect subsequent choices. None is guaranteed. Political identity may matter little relative to model quality, price, or convenience; heterogeneous feedback may wash out during alignment; induced changes may remain narrowly confined to the training examples; or providers may suppress divergence through balanced sampling, system-level constraints, pooling, and regularization. We provide evidence concerning the first two links and use an agent-based model to examine what follows when model choice and adaptation are allowed to operate together over time. Figure~\ref{fig} summarizes the proposed mechanism and how each of its components is studied.

We study the process in three steps. First, we draw on a previously reported human model-choice experiment that provides evidence for political sorting across AI systems \citep{HeseltineChoice2026}. Political identity predicted which system participants selected, including when they were paid for accurate answers: Republicans were more likely than Democrats to choose Grok, Democrats were more likely to choose Claude, and 71\% of participants returned to a model they had previously preferred. These findings show that competing AI systems can attract politically different audiences even when model choice has instrumental consequences.

Second, we test whether politically structured feedback can cause initially identical models to behave differently. We fine-tune paired copies of Qwen2.5-1.5B, Mistral-7B, and GPT-OSS-20B using predominantly Democratic or predominantly Republican synthetic feedback derived from OpinionQA survey distributions \citep{santurkar2023whose}. The deliberately strong 90/10 audience contrast is a stress test of whether audience composition can generate broader behavioral differentiation, rather than an estimate of present-day sorting in AI markets. Across five independent runs per model family, paired models selected different answers on 12--41\% of unseen questions with large partisan gaps, and the average difference followed the political ordering of the feedback in every run. For some models, differentiation also persisted in political domains excluded from preference training. Audience-specific feedback can therefore affect behavior beyond the precise questions on which it was supplied, although transfer varies across models, domains, and prompts. An exploratory identity-cue experiment additionally asks whether sycophantic personalization can reduce between-model differentiation: rather than shifting a model globally toward one audience, alignment may lead the same model to answer Democrats and Republicans differently when their identities are explicit, accommodating political heterogeneity within a single system. We treat this as a complementary pathway that could weaken sorting across models while creating differentiation within them.

Third, we ask what happens if political sorting and feedback-based alignment are coupled over time. The two links can be observed separately in experiments, but their joint consequence is not the effect of any single update: it is the trajectory produced by repeated cycles of choice, feedback, and adaptation, which no one-shot experiment can observe. We therefore introduce a framework that brings LLM fine-tuning into agent-based modeling. Recent work has used LLMs to simulate human agents and enrich behavior within social simulations \citep{Aher2023Turing,Park2023Generative,Gao2024LLMABM,Piao2025AgentSociety}. We take a different approach: rather than using an LLM to stand in for the human agent, we use fine-tuning experiments to measure how the AI system itself changes in response to socially structured feedback, and embed that measured adaptation in an ABM of repeated human--AI interaction. Model adaptation is anchored in the fine-tuning experiments and user choice in the previously reported human experiment. Initially small model differences shape which audiences models attract; politically distinct audiences generate different feedback streams; and models that learn from those streams become more different still. Pooling feedback breaks the link between audience composition and model adaptation, while regularization dampens the resulting divergence.

Our contribution is both substantive and methodological. Substantively, we identify the \emph{centrifugal alignment spiral} as a mechanism through which political sorting can drive AI models apart: users do not merely select among fixed information sources, but their choices can shape the systems available for future choice. Methodologically, we connect experimental LLM fine-tuning to agent-based modeling, providing a framework for studying sociotechnical systems in which human and AI behavior coevolve. Our claim is not that fragmentation is inevitable---commercial systems can filter, balance, pool, or regularize feedback, and our experiments show that such choices matter. The broader implication is that alignment data have a social origin: when different audiences help train different systems, political sorting can become a mechanism of durable model differentiation.

\section*{Results}

\subsection*{Politically sorted feedback pushes AI models apart}

We first test the alignment link in the centrifugal alignment spiral: can feedback from politically different audiences cause initially identical models to behave differently? We use OpinionQA, a dataset constructed from Pew Research Center's American Trends Panel surveys that contains multiple-choice questions and respondent-level answers, allowing response distributions to be calculated separately for Democrats and Republicans \citep{santurkar2023whose}. We fine-tune two copies of the same base model in each run. One copy receives synthetic feedback drawn from an audience composed of 90\% Democrats and 10\% Republicans (Dem90), while the other receives the reverse composition (Rep90). This deliberately strong contrast is designed to test whether audience composition can generate broader model differentiation, rather than to reproduce current levels of political sorting in AI markets. Within each run, the paired models use the same training questions and candidate responses; only the political composition of the feedback differs.

We evaluate the fine-tuned models on 160 OpinionQA questions from a disjoint test set. These questions have large differences between Democratic and Republican response distributions and were used neither to generate candidate responses nor to construct preference pairs. We examine two outcomes. \emph{Choice divergence} measures the proportion of questions on which the Dem90 and Rep90 models select different answers. The \emph{directional political gap} measures whether the Dem90 model selects answers with greater relative support among Democrats than the corresponding answers selected by the Rep90 model.

\begin{figure*}[t]
\centering
\includegraphics[width=\textwidth]{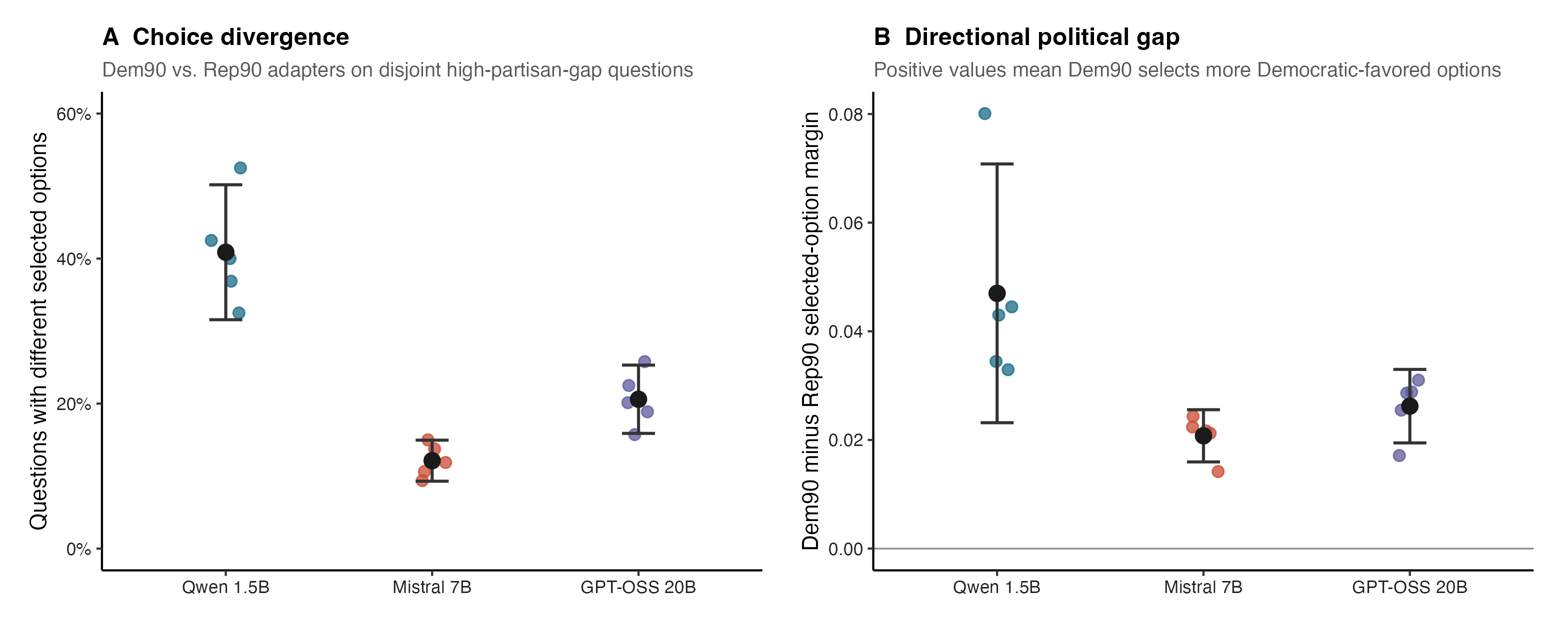}
\caption{\textbf{Politically sorted feedback differentiates paired AI models.} Democratic- and Republican-trained copies of the same base model are evaluated on 160 high-partisan-gap OpinionQA questions that entered neither candidate generation nor preference-pair construction. Points represent independent end-to-end training runs; bars show 95\% $t$ intervals across the five runs. Choice divergence is the proportion of questions on which the paired models select different answers. The directional political gap measures whether the Dem90 model selects answers with greater Democratic-relative-to-Republican support than the corresponding Rep90 model.}
\label{fig:generalization}
\end{figure*}

Politically sorted feedback differentiated all three model families (Fig.~\ref{fig:generalization}). Across five independent end-to-end runs, paired Qwen2.5-1.5B models selected different answers on an average of 40.9\% of the unseen high-partisan-gap questions. The corresponding averages were 12.1\% for Mistral-7B and 20.6\% for GPT-OSS-20B. The differences were also politically ordered: in every run for all three model families, the Dem90 model selected answers that were, on average, more Democratic-favored than those selected by the corresponding Rep90 model (one-sided sign test across the five runs, $p=0.031$ for each family).

The magnitude and form of differentiation varied across model families. Qwen showed the largest response to the feedback manipulation, Mistral the smallest, and GPT-OSS an intermediate response. The effect was also partly sensitive to prompting. Under an alternate instruction and response format, choice divergence persisted, although the directional political gap weakened for GPT-OSS. The adaptation was therefore probabilistic rather than uniform: individual answer changes did not always follow the political direction of the feedback. Nevertheless, the average directional effect was consistent across independent runs.

A matched training-intensity comparison reported in the SI provides an additional test. Under a lower-intensity alignment regime, paired models diverged on 3.1\% of unseen high-partisan-gap questions for Qwen and 3.8\% for Mistral. Under the stronger regime used in the principal analysis, divergence increased substantially. Stronger preference-based updating therefore produced greater behavioral separation between models trained on politically different feedback.

Because the design does not include paired models trained independently on the same feedback composition, choice divergence may include some separation caused by optimization stochasticity. The consistent political ordering across runs, however, cannot be explained by undirected training variation alone and provides the clearest evidence that feedback composition systematically shaped model behavior.

\subsection*{Political differentiation extends beyond the domains used for alignment}

Performance on unseen questions shows that the models are not simply reproducing answers encountered during training. However, test questions may still resemble training questions from the same substantive issue area. We therefore conduct a stricter cross-domain test, by  dividing the political questions into four broad domains: foreign policy; government, democracy, and media; economic policy and welfare; and culture, identity, and immigration. For each domain, we remove all questions from that domain before candidate generation and preference construction, fine-tune the Democratic- and Republican-trained models on the remaining domains, and evaluate them only on questions from the excluded domain. Each model-by-domain condition is repeated across five independent training runs and two neutral prompt formats. Each question is presented independently, without information about the user’s political identity or any conversational history. The observed differences therefore reflect changes in the models’ response behavior rather than adherence to an explicitly supplied persona.

\begin{figure*}[t]
\centering
\includegraphics[width=\textwidth]{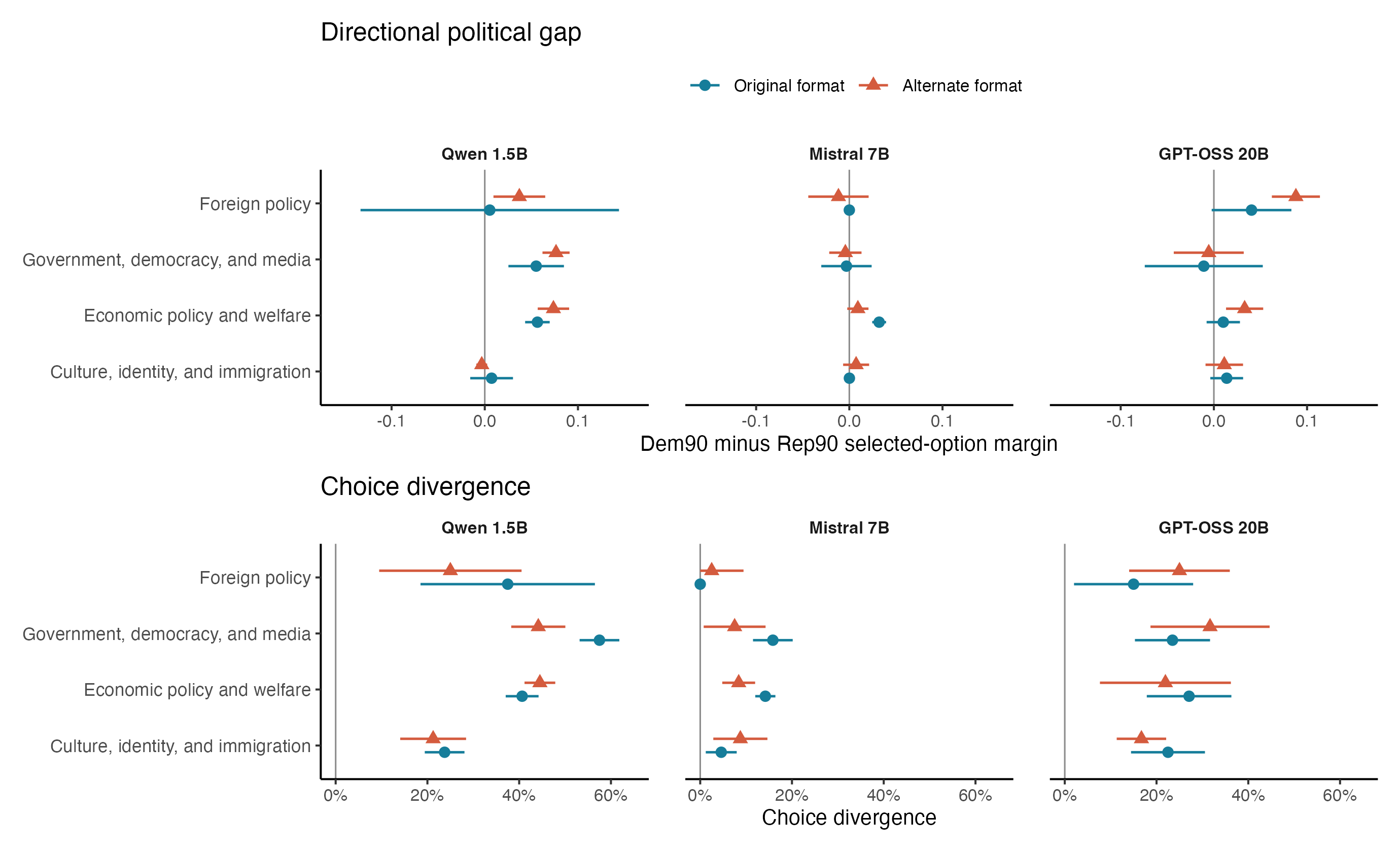}
\caption{\textbf{Model differentiation can extend to issue domains withheld from alignment.} For each condition, all training questions from the indicated substantive issue domain are removed before candidate generation and preference construction. Democratic- and Republican-trained models are then evaluated only on questions from that excluded domain. Points show means across five independent training runs; bars show 95\% $t$ intervals across runs. Choice divergence measures how often the paired models select different answers, whereas the directional political gap measures whether those differences follow the Democratic--Republican ordering of the training feedback. Valid evaluated-item counts are 8 for foreign policy, 23--24 for government, democracy, and media, 31 for economic policy and welfare, and 47--48 for culture, identity, and immigration, depending on model and prompt format. Cross-domain transfer varies across model families, issue domains, and prompt formats.}
\label{fig:cross-domain}
\end{figure*}

Qwen retained substantial differentiation even when it received no preference feedback from the domain on which it was evaluated (Fig.~\ref{fig:cross-domain}). Averaging equally across the four excluded domains, paired models selected different answers on 39.8\% of questions under the original prompt format and 33.7\% under the alternate format. These choice-divergence rates capture all differences between the paired models, regardless of political direction. The directional political gap asks a narrower question: whether the Qwen model trained on predominantly Democratic feedback selected answers with greater Democratic-relative-to-Republican support than the model trained on predominantly Republican feedback. This gap remained positive in every run under both prompt formats. Compared with models trained on questions from all four domains, Qwen retained approximately two-thirds of the original directional gap under one prompt format and more than four-fifths under the other.

Cross-domain transfer was weaker and less consistent for the other model families. Mistral showed stable politically directional transfer mainly on economic-policy and welfare questions. GPT-OSS showed intermediate levels of choice divergence, but the political direction of those differences varied more across issue domains and prompt formats. Thus, paired models could behave differently without those differences always corresponding to a consistent Democratic--Republican ordering. 

Politically structured feedback did not produce a single, prompt-invariant ideological shift across all models and domains. Its effects were nevertheless not confined to the subjects represented during alignment. For Qwen, and more selectively for Mistral and GPT-OSS, feedback from politically different audiences changed responses in substantive issue areas that had been entirely excluded from training. This suggests that fine-tuning can induce broader associations among partisan preferences and political positions, allowing differentiation to extend beyond the topics that initially generated the feedback. The unevenness across models and domains, however, indicates that this process produces structured but model-specific patterns of differentiation rather than a coherent ideological shift.

\subsection*{Accumulation and pooling shape model differentiation}

The centrifugal alignment spiral depends not only on whether feedback can change a model, but also on what happens as feedback accumulates and how providers distribute it across systems. Repeated feedback from politically distinct audiences could gradually amplify model differences. One possible response, for providers that use audience feedback in alignment, would be to pool feedback across users rather than allow each model to learn primarily from its own audience. We use a diagnostic experiment to test both possibilities.

The experiment uses Qwen2.5-0.5B, whose smaller size made it feasible to vary feedback volume and pooling systematically. Because each condition was run once with a single model family, the results are intended to clarify the proposed mechanisms rather than estimate effects that generalize across architectures.

We first ask whether political differentiation grows as models receive more feedback from the same audience. We hold both the training questions and audience composition fixed while increasing the number of synthetic users contributing preference signals to each question. As the number of users increases from 10 to 20 to 50, choice divergence on unseen questions rises from 20.6\% to 26.9\% to 32.9\%. Thus, behavioral separation can accumulate even when audiences provide feedback on the same questions and do not express increasingly extreme preferences.

We next ask whether pooling feedback across political groups could suppress this accumulation. We progressively replace feedback drawn from each model's predominantly Democratic or Republican audience with feedback drawn from the overall OpinionQA population. Choice divergence declines monotonically from 31.9\% with fully separate feedback streams to 21.3\% with 25\% pooling, 16.9\% with 50\% pooling, 10.0\% with 75\% pooling, and 3.8\% under full pooling. The directional political gap falls sharply overall, from 0.057 under separate feedback to 0.009 under full pooling, with a small uptick at the final pooling step. These results suggest that pooling feedback across political groups may provide a design lever for suppressing between-model differentiation.

\subsection*{Personalization can shift political differentiation within models}

The preceding experiments examine one way that politically different audiences may shape AI systems: separate models may shift globally toward the preferences of their own users. A model could also adapt in a different way: rather than changing its general political orientation, it might learn to tailor its answers to what it knows about the current user. A single model could then give more Democratic-aligned responses to a user identified as a Democrat and more Republican-aligned responses to a user identified as a Republican. Such sycophantic personalization could, in principle, accommodate political heterogeneity within one shared system and reduce pressure for models to fragment, while still producing political differentiation across individual interactions.

We examine this possibility in an exploratory identity-cue experiment with one training run per model family. Each model is trained on a balanced mixture of Democratic- and Republican-aligned preferences, and every preference example explicitly identifies the user as a Democrat or Republican. The model can therefore learn to condition its responses on user identity rather than shifting globally toward either group. We evaluate neutral, Democratic-cued, and Republican-cued versions of 160 disjoint test questions. Conditional adaptation is measured by how often the two identity cues produce different answers and by the difference in the political margins of those answers.

The extent of conditional adaptation varied substantially across models (SI Appendix). For Mistral-7B, Democratic and Republican identity cues produced different answers on approximately 19--24\% of disjoint test questions, with Democratic-minus-Republican shifts in political margin of about 0.05--0.08. Qwen2.5-1.5B was much less responsive: the cues changed answers on approximately 3--4\% of questions, and directional shifts were close to zero. GPT-OSS showed cue effects that were less consistent across prompt formats.

These results distinguish two possible forms of political adaptation. Models may shift globally toward the audience supplying their feedback, producing differentiation between systems, or they may adapt conditionally to individual users within a shared system. Conditional adaptation could weaken between-model fragmentation by serving heterogeneous audiences within one model, but it could also reproduce political separation less visibly through personalized responses.

\subsection*{Coupling audience sorting and feedback learning can make model differences self-reinforcing}

The fine-tuning experiments show that audience composition can systematically alter model behavior. The previously reported human experiment provides the complementary sorting link: political identity predicts model choice, including when participants are rewarded for accuracy \citep{HeseltineChoice2026}. We next use an agent-based model to ask what happens when model choice and feedback-based adaptation repeatedly influence one another over time.

The model contains 5,000 politically heterogeneous users and four initially similar AI models. At each time step, users choose among models according to political fit. Each model then updates toward the mean feedback supplied by its current audience, while regularization pulls all models toward a shared baseline. In targeted scenarios, we also introduce an empirically anchored tendency to return to the previously selected model. Choice sensitivity is anchored in the human experiment, and model responsiveness is varied over the range implied by the fine-tuning experiments. The purpose of the ABM is to isolate the dynamic consequences of coupling the two empirically observed links. Full equations and anchoring procedures are reported in Materials and Methods and the SI Appendix.

\begin{figure*}[p]
\centering
\includegraphics[width=\textwidth]{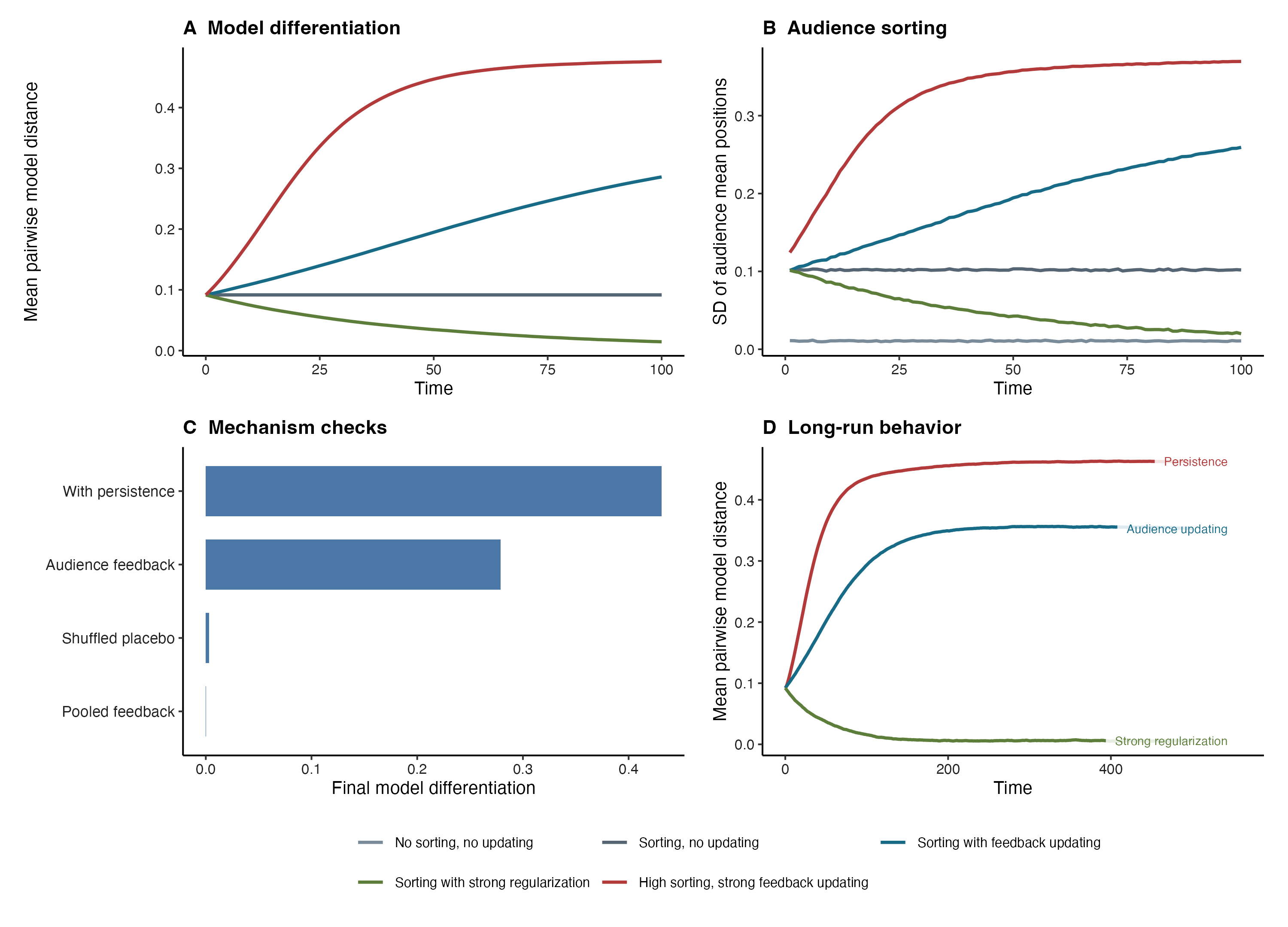}
\caption{\textbf{Coupling audience sorting to model adaptation produces the centrifugal alignment spiral.} 
(A) Political differentiation among models over time. 
(B) Political segregation among model audiences. 
(C) Mechanism checks: audience-specific updating increases model differentiation, whereas shuffled feedback, pooled feedback, and stronger regularization suppress it; persistence in model choice strengthens it. 
(D) Long-run dynamics under audience-specific updating and stronger regularization.}
\label{fig:main}
\end{figure*}

When audience sorting and model adaptation are coupled, they produce the centrifugal alignment spiral (Fig.~\ref{fig:main}). Small initial differences among models affect the political composition of the audiences they attract. Those audiences supply systematically different feedback, moving the models farther apart. The resulting model differences then shape the next round of user choices. Model differentiation and audience segregation consequently reinforce one another.

The process is strongly path-dependent. Across simulations, initial and final model positions correlate at 0.974, with a mean rank correlation of 0.927. Models that begin slightly to the left or right generally retain their relative ordering as audience-specific feedback increases their separation. Within the model, durable political differentiation can therefore emerge from small initial perceived differences without any provider intentionally assigning models distinct political positions.

Mechanism checks confirm that this result depends on models learning from their own audiences. In the shuffled-feedback condition, users choose models as usual, but their feedback is randomly reassigned before model updates. This preserves both the pattern of user choices and the overall distribution of feedback while breaking the association between each model and its audience. Under this condition, model differentiation is largely eliminated. Giving every model the same pooled feedback target similarly suppresses divergence. Stronger regularization dampens differentiation, whereas a tendency to return to the same model strengthens it. These comparisons show that sorting becomes self-reinforcing only when it generates distinct feedback streams that remain connected to the models whose audiences produced them.

The spiral is not inevitable: parameter sweeps reveal a transition region in which final model differentiation begins to exceed initial differentiation. Amplification occurs when users are sufficiently responsive to political fit and models are sufficiently responsive to audience feedback relative to the strength of regularization. Greater model responsiveness lowers the degree of audience sorting required for amplification, whereas stronger regularization raises it. Persistence in model choice can move otherwise marginal systems across this boundary. Across the range of responsiveness implied by the fine-tuning experiments, the human choice results generate plausible scenarios on both sides of the transition, depending on how strongly users perceive political differences among models.

These results also identify practical design levers. Routing feedback separately by model audience makes differentiation more likely, whereas pooling or balancing feedback across audiences and more strongly regularizing updates make it less likely. These results suggest concrete steps that developers could evaluate if political fragmentation becomes a concern.

The ABM therefore does not predict that AI markets \emph{must} fragment. It identifies the conditions under which the separately observed processes of audience sorting and feedback learning can become mutually reinforcing. When political sorting produces distinct feedback streams and models remain responsive to those streams, small initial differences can develop into persistent separation between both models and their audiences.

\section*{Discussion}

The social-media era made the fragmentation of political information environments a central concern in political communication. Debates over audience sorting, echo chambers, misinformation, and polarization have increasingly centered on whether digital media undermine a shared epistemic foundation. This study identifies a corresponding risk in an adaptive AI ecosystem: fragmented audiences may not only inhabit different information environments, but help make those environments different through a feedback process. Political differences shape which models users choose; those choices determine which audiences supply feedback to which systems; and audience-specific alignment can push those systems farther apart. Once coupled, model differentiation and audience sorting can become mutually reinforcing. We call this process the \emph{centrifugal alignment spiral}. 

Our results provide evidence for the two empirical links required by this mechanism and model the consequences of coupling them. A previously reported human experiment shows that political identity predicts model choice, including when participants are rewarded for accuracy. Our fine-tuning experiments show that audience composition can, in turn, become a cause of model behavior. Copies of the same base model trained on predominantly Democratic or Republican feedback diverged on unseen political questions, with politically ordered differences in every replicated run across three model families. Diagnostic analyses further suggest that repeated audience-specific feedback increases separation, whereas drawing feedback from a shared population suppresses it. When the sorting and adaptation links are coupled in the agent-based model, small initial differences can grow as models attract distinct audiences and learn from them.

A central question is how far feedback-induced adaptation extends: if models simply reproduce preferences expressed on the training questions, the potential for broader fragmentation is limited. We find that differentiation persists on questions that entered neither candidate generation nor preference construction. For some models, it also extends to issue domains entirely excluded from alignment. Politically structured feedback can therefore alter responses beyond the topics that originally produced it. This transfer is nevertheless uneven: its magnitude and political direction vary across models, issue domains, and prompts.

This heterogeneity places an important boundary on the argument. The centrifugal alignment spiral does not require models to acquire a coherent ideology resembling that of a human political actor. It requires only that different feedback streams produce persistent behavioral differences that matter to future users. Those differences may be domain-specific, prompt-sensitive, or organized differently across architectures. The variation we observe suggests that political adaptation may operate through model-specific associations among issues rather than movement along a single ideological dimension. Our one-dimensional ABM deliberately abstracts from this structure to isolate the population-level feedback mechanism. Future models should examine multidimensional political spaces in which adaptation within one domain spills over unevenly into others.

The mechanism extends familiar concerns about feedback loops in digital systems \citep{Chaney2018,Jiang2019,Pagan2023FeedbackLoops}. In recommender systems, user behavior shapes which content is subsequently ranked and displayed. In adaptive AI systems, feedback can alter the response policy of the information source itself. Political sorting can therefore move from the demand side of the information environment into the production of the source. This process may also be difficult to observe. Users generally interact with models privately, while providers disclose little about how feedback is routed or incorporated into later updates. Model differentiation could consequently emerge without being immediately visible either to individual users or to the organizations operating the systems.

Personalization offers another possible response to political heterogeneity. Rather than shifting globally toward one audience, a shared model may learn to tailor its answers to the user in front of it. Our exploratory identity-cue experiment suggests that this capacity varies across models: cue effects were substantial for Mistral, weak for Qwen, and less stable across prompt formats for GPT-OSS. This creates a suggestive substitution pattern: Mistral showed the weakest between-model differentiation in the principal experiment but the strongest within-model identity-cued response, whereas Qwen showed the reverse. Personalization could reduce pressure for separate models to diverge by allowing one system to serve politically different audiences. It could also create a less visible form of fragmentation if users first choose congenial models and then receive further politically tailored responses within them. Its broader consequences will depend on whether personalization reduces sorting across systems or deepens differentiation within already separated audiences; the one-run analysis is exploratory rather than a test of that ecosystem-level tradeoff.

Our findings moreover identify several possible interventions against model fragmentation. If competing models learn primarily from their own audiences, political sorting can generate increasingly different alignment data and, eventually, increasingly different systems. Providers could weaken this link by pooling or balancing feedback before model updates. In our diagnostic experiment, repeated audience-specific feedback increased behavioral separation even without introducing new topics or more extreme preferences, whereas feedback drawn from a shared population largely suppressed divergence. The ABM isolates the same mechanism: when models receive a common feedback target, audience sorting no longer produces model-specific adaptation. Balanced sampling, stronger regularization, and regression tests across alternative audience compositions therefore offer concrete safeguards that developers could evaluate.

Yet interventions within a single company may be insufficient if different providers attract politically distinct publics. Preventing ecosystem-wide fragmentation could require some coordination across firms, for example through shared reference datasets, common evaluation standards, independent audits, or public-interest feedback infrastructures. Such cooperation may be difficult to sustain voluntarily: companies have incentives to protect proprietary data, differentiate their products, and cater to their existing users. Policymakers may therefore have a role in creating standards and institutions that make coordination possible without requiring companies to exchange raw user data or compromise privacy and competition.

This raises a deeper democratic question. Pooling feedback does not remove politics from alignment; it shifts the question to whose preferences define the shared target. Any common baseline must decide which groups are represented, how disagreement is aggregated, which forms of pluralism should be preserved, and when minority claims should be protected against majority preferences. These decisions should not be left implicitly to market share, engineering convenience, or the composition of whichever users provide the most feedback. The central challenge is to build democratic scaffolding around alignment: transparent procedures, independent oversight, meaningful public participation, and mechanisms for contesting the political and epistemic assumptions embedded in widely used systems. The political character of AI systems is likely to become a defining political issue and should be subject to democratic deliberation, oversight, and contestation, rather than left to emerge from opaque feedback pipelines, proprietary alignment choices, and the accidental composition of user populations.

The paper also makes a methodological contribution. Existing LLM-based social simulations have largely used language models to generate or enrich the behavior of simulated human agents \citep{Park2023Generative,tornberg2023simulating}. We instead use fine-tuning experiments to make the AI system itself an empirically anchored adaptive component of the social model. This allows human behavior to change AI behavior and those changing systems, in turn, to reshape human choice. Many important consequences of AI will arise from neither humans nor models in isolation, but from their repeated interaction. Combining experimental model adaptation with agent-based simulation provides one way to study these emerging sociotechnical dynamics.

Several limitations define the scope of our findings. This study is designed as a stylized test of a proposed mechanism, rather than a reconstruction of current AI markets or commercial alignment pipelines. Its empirical components isolate two links: whether politically different audiences select different models, and whether politically structured feedback can make initially identical models behave differently. OpinionQA provides a transparent setting in which audience composition, preference signals, and political direction can be controlled and measured. The 90/10 audience contrast, group-level preference construction, high-partisan-gap evaluation items, and stronger DPO regime are therefore intended to make the mechanism observable, not to estimate its current magnitude in deployed systems. This design abstracts from important features of real-world use: synthetic feedback does not capture the full heterogeneity of user interactions, multiple-choice attitudes represent only one form of political behavior, and LoRA-based DPO on open models differs from production alignment. Because OpinionQA supplies both the feedback distributions and the directional outcome measure, the principal analysis tests generalization across questions rather than across data sources; the cross-domain analyses provide a stronger transfer test, but their results vary across architectures, issue areas, and prompts. 
One link in the proposed spiral also remains untested directly: the companion experiment shows that political identity predicts choices among existing AI systems, but we do not show that users sort in response to the specific differences produced by the Dem90- and Rep90-trained models. A blinded choice experiment between these adapters would close that loop. 
Our study therefore identifies a plausible mechanism and the conditions under which it can become self-reinforcing; establishing its prevalence and magnitude will require direct evidence from deployed feedback pipelines, model updates, and user behavior.

\section*{Materials and Methods}

\begingroup
\small

\subsection*{Study design}

The study combines two empirical components with an agent-based model. First, we draw on previously reported evidence from a human model-choice experiment to characterize political sorting across AI systems \citep{HeseltineChoice2026}. Second, we conduct feedback-based alignment experiments to estimate how strongly audience composition can differentiate initially identical language models and whether these differences extend beyond the questions used for alignment. Third, the ABM asks what follows if these separately observed processes feed back into one another over time.

The human experiment is reported elsewhere and is not reanalyzed as a new experiment in this paper. We use its reported model-choice and persistence estimates to anchor plausible parameter values in the ABM. The language-model experiments and ABM simulations are original to this study.

\subsection*{Human model choice and the sorting link}

We use reported results from a companion experiment in which 1,884 U.S. participants evaluated outputs from four AI systems and subsequently selected a model under accuracy incentives \citep{HeseltineChoice2026}. Participants were randomly assigned in a 2-by-2 design varying recognizable brand names versus neutral labels and political-first versus general-knowledge-first exposure. The experiment provides two quantities relevant to the ABM: partisan differences in model choice and persistence in previous model choice.

In the ABM, the parameter $\alpha$ governs how strongly users respond to political fit between themselves and a model. The human experiment does not identify $\alpha$ directly because the same observed choice difference can arise from strong sensitivity to subtle perceived model differences or weaker sensitivity to larger differences. We therefore map the observed choice gaps onto plausible values of $\alpha$ under alternative assumptions about perceived political distance.

In a symmetric binary-choice approximation, let $2d$ denote the difference in ideological distance between the politically more congenial and less congenial alternatives. The probability of choosing the congenial model is

\begin{equation}
P(\mathrm{congenial})=
\operatorname{logit}^{-1}(2\alpha d),
\end{equation}

and the corresponding between-group choice gap is

\begin{equation}
g=2\operatorname{logit}^{-1}(2\alpha d)-1=\tanh(\alpha d)
\end{equation}

Thus,

\begin{equation}
\alpha \approx \frac{\operatorname{atanh}(g)}{d}.
\end{equation}

The 10.3-percentage-point Democrat--Republican choice gap observed in the initial task corresponds to $\alpha=1.03$ when $d=0.10$ and $\alpha=2.07$ when $d=0.05$. The 19.9-point gap observed in the later incentivized task corresponds to $\alpha=2.02$ and $\alpha=4.03$, respectively. These calculations define plausible scenarios rather than point estimates of a structural choice parameter.

The values $d=0.05$ and $d=0.10$ are sensitivity benchmarks, not independently estimated perceptual distances. They correspond to cases in which the politically relevant difference between two model outputs is small on the $[-1,1]$ ideological scale but still large enough to be noticed by users. We use them because the companion experiment shows sizable choice differences among models whose substantive differences are likely modest. Broader sensitivity analyses use perceived-distance scenarios from 0.05 to 0.40 and ABM choice-sensitivity values from $\alpha=0$ to $\alpha=5$.

Persistence is anchored separately. Let $\rho$ denote the utility bonus associated with returning to the model selected in the preceding period. Under an equal-baseline approximation with $M$ models, the persistence bonus associated with a repeat-choice probability $p$ is

\begin{equation}
\rho=
\log\left(
\frac{(M-1)p}{1-p}
\right).
\end{equation}

With four models, the reported pooled 71\% repeat-choice rate implies $\rho=1.99$. The corresponding estimates are $\rho=1.72$ and $\rho=2.31$ in the unnamed- and named-model conditions, respectively. Repeat choice is not a structural estimate of inertia: it can also reflect stable preferences, brand affinity, perceived quality, or ideological fit. The benchmark ABM reported in the main comparison therefore sets $\rho=0$ so that amplification can be attributed to political sorting and audience-specific feedback alone. We then use $\rho=1.99$ in a targeted persistence scenario, where it should be read as an empirically anchored repeat-choice tendency rather than a pure switching-cost parameter.

These mappings are used only to anchor plausible regions of the ABM parameter space. They do not assume that political fit is the sole determinant of model adoption or that the experimental choice setting reproduces real-world model markets.

\subsection*{OpinionQA data}

We use OpinionQA, a dataset derived from Pew Research Center’s American Trends Panel surveys that includes multiple-choice questions and respondent-level answers, enabling response distributions to be calculated separately for Democrats and Republicans \citep{santurkar2023whose}. The frozen processed corpus contains 1,078 questions in the training split, 231 in a validation split, and 231 in a disjoint test split.

The principal fine-tuning experiments use 500 politically diagnostic questions from the training split. Evaluation is conducted on 160 questions from the disjoint test split with the largest Democratic--Republican differences in answer distributions. These test questions enter neither candidate generation nor preference-pair construction.

The 500 training questions are the 500 training-split items with the largest Democratic--Republican distributional gaps among the 1,078 available training items. The gap is the total-variation distance between Democratic and Republican answer-option distributions, $\frac{1}{2}\sum_a |s_D(a_q)-s_R(a_q)|$. For diagnostic in-sample evaluation, we also evaluate models on the top 160 training-split items by the same criterion. This 160-item in-sample evaluation set is a subset of the 500-question DPO pool and is reported only to compare in-sample and disjoint-test behavior.

For each answer option $a$ to question $q$, OpinionQA provides Democratic and Republican support, denoted $s_D(a_q)$ and $s_R(a_q)$. We use these group-specific distributions both to construct synthetic feedback and to evaluate the political direction of model responses.

\subsection*{Generating politically structured feedback}

For each training question, the base model generates eight candidate responses. Candidate generation uses sampling with temperature 0.9, top-$p=0.95$, and a 60-token generation limit. Prompts require the response to select exactly one of the listed answer options and begin with \texttt{Selected option: X}. We parse the selected option from each response and associate it with the Democratic and Republican support for that option in OpinionQA.

Synthetic users are then sampled from the specified audience composition. In the principal contrast, one feedback stream contains 90\% Democratic and 10\% Republican synthetic users (Dem90), whereas the other contains 10\% Democratic and 90\% Republican users (Rep90). Conditional on group identity, the synthetic user ranks the available valid candidates by that group's OpinionQA support for the selected option. This uses OpinionQA distributions to construct group-level preference rankings, not to sample heterogeneous individual preferences from the full within-party distribution.

For the generated-candidate DPO experiments, the synthetic user's group is sampled from the specified audience mixture. Conditional on that group, each valid candidate is scored by the OpinionQA support for its parsed answer option. The preferred response is sampled uniformly from the candidates tied for the highest group-specific support, and the rejected response is sampled uniformly from the candidates tied for the lowest group-specific support. Invalid or unparsable candidates are excluded before scoring. Duplicate candidates and duplicate selected options are retained in the candidate pool; ties are resolved by random sampling within the tied best or worst pool. If fewer than two valid candidates remain, or if all valid candidates have the same group-specific support, the synthetic user contributes no pair for that question. Otherwise, each synthetic user contributes one DPO pair per question, and duplicate preference pairs are retained as repeated training observations. This design intentionally creates a strong group-level feedback signal; probabilistic pairwise sampling that preserves within-party heterogeneity is an important robustness extension.

Dem90 and Rep90 adapters within the same run always begin from the same base model and use the same training questions and candidate-response pool. The intended contrast is the political composition of the synthetic feedback. This paired design removes stochastic differences in candidate generation, but it does not include same-composition training controls; therefore, directional political ordering is the strongest evidence for composition-specific movement, while raw choice divergence should be interpreted as behavioral separation under this paired training protocol.

\subsection*{Feedback-based alignment}

The principal experiments use Qwen2.5-1.5B-Instruct, Mistral-7B-Instruct-v0.3, and GPT-OSS-20B. We train separate low-rank adaptation (LoRA) adapters using direct preference optimization (DPO).

For Qwen and Mistral, we compare two alignment intensities. The conservative regime uses one epoch and a learning rate of $10^{-6}$. The stronger regime uses two epochs and a learning rate of $3\times10^{-6}$. The candidate pool, synthetic-user count, feedback composition, and random seed are held fixed within the matched comparison. GPT-OSS is evaluated under the stronger regime as an architecturally distinct and larger-model replication.

The principal experiments use 50 synthetic users per training item. Within every run, Dem90 and Rep90 adapters receive feedback generated from the same question and candidate pools. We repeat the stronger-training experiment across five independent end-to-end runs. In each run, candidate generation, synthetic preference sampling, and DPO optimization are repeated from scratch while the processed train--test split remains fixed.

The five principal replication seeds are 202, 203, 204, 205, and 206. For Qwen and Mistral, LoRA adapters use rank 8, LoRA alpha 16, dropout 0.05, and target modules \texttt{q\_proj} and \texttt{v\_proj}. DPO uses $\beta=0.1$, per-device batch size 1, gradient accumulation 16, maximum sequence length 1,024, maximum gradient norm 0.3, gradient checkpointing, the TRL default AdamW optimizer and learning-rate schedule, and no explicit warmup. Qwen and Mistral runs use fp32 precision. GPT-OSS uses the same LoRA and DPO settings but bf16 precision and longer generation limits. The main Qwen, Mistral, and GPT-OSS runs were executed on the Snellius GPU partition using two H100-class GPUs per paired run; the smaller Qwen2.5-0.5B accumulation and pooling experiments were run on a two-GPU A10 server. The reproducibility environment used Python 3.11 on Snellius; the server requirements pin PyTorch 2.6.0 with CUDA 12.4 for the non-GPT-OSS runs, and the GPT-OSS environment is described in the replication archive.

GPT-OSS uses its native Harmony chat template with low reasoning effort. Evaluation extracts the final response channel rather than text from the model's reasoning channel.

\subsection*{Evaluation of model differentiation}

Evaluation is deterministic, with no sampling and an 80-token generation limit. The principal prompts contain no information about the political identity of the user. We additionally evaluate each model under an alternate instruction and response format while holding the substantive question wording and answer options unchanged. Evaluation responses are included in outcome calculations only when the selected option can be parsed and corresponds to a listed answer option; invalid paired responses are excluded from pairwise contrasts.

We use two complementary outcomes. \emph{Choice divergence} is the proportion of questions on which the paired Dem90 and Rep90 adapters select different answer options. It captures any behavioral separation, regardless of political direction.

To measure whether separation follows the political ordering of the feedback, define the political margin of a selected answer $a_q$ as

\begin{equation}
m(a_q)=s_D(a_q)-s_R(a_q).
\end{equation}

For question $q$, directional separation between the paired adapters is

\begin{equation}
\begin{aligned}
\Delta_q={}&s_D(a_q^{\mathrm{Dem90}})-s_R(a_q^{\mathrm{Dem90}})\\
&-s_D(a_q^{\mathrm{Rep90}})+s_R(a_q^{\mathrm{Rep90}}).
\end{aligned}
\end{equation}

or equivalently,

\begin{equation}
\Delta_q = m\left(a_q^{\mathrm{Dem90}}\right) - m\left(a_q^{\mathrm{Rep90}}\right).
\end{equation}

Positive values indicate that the Dem90 adapter selected an answer with greater Democratic-relative-to-Republican support than the answer selected by the Rep90 adapter. We refer to the mean of $\Delta_q$ across questions as the directional political gap.

Across-run summaries report means and standard deviations over independent training runs. Intervals in the main figures are $t$ intervals over independent runs when a run-level replication exists. Single-run diagnostic analyses, such as some exploratory issue-level summaries, use item-level normal approximations only to describe variation across evaluated questions within that run. We do not interpret item-level intervals as generalizing over training randomness.

\subsection*{Cross-domain holdout experiments}

Evaluation on a disjoint test split rules out direct question-level memorization but cannot determine whether adaptation transfers beyond political domains represented in training. We therefore conduct a stronger cross-domain holdout experiment.

We define four sufficiently large political domains: foreign policy; government, democracy, and media; economic policy and welfare; and culture, identity, and immigration. For each experiment, all training questions assigned to the focal domain are removed \emph{before} candidate generation and preference-pair construction. Separate Dem90 and Rep90 adapters are then trained on 500 questions from the remaining domains and evaluated only on test questions assigned to the excluded domain.

Each model--domain condition is repeated for seeds 202--206. Models train for two epochs at a learning rate of $3\times10^{-6}$ using eight candidates and 50 synthetic users per training question. Evaluation uses both the original and alternate neutral prompt formats. The excluded domains contain between 8 and 48 valid test questions. Intervals in Fig.~\ref{fig:cross-domain} are $t$ intervals across the five independent training runs; the corresponding valid item counts are 8 (foreign policy), 23--24 (government, democracy, and media), 31 (economic policy and welfare), and 47--48 (culture, identity, and immigration), depending on model and prompt format.

The focal domains contain 43 foreign-policy, 109 government/media, 177 economic-policy, and 279 culture/identity/immigration questions in the training split. Removing them leaves 1,035, 969, 901, and 799 training questions, respectively. Because each remaining pool exceeds 500 items, every cross-domain run uses the 500 remaining training questions with the largest partisan gaps. The held-in training-question set is fixed across seeds within each domain; seeds vary candidate generation, synthetic preference sampling, and DPO optimization.

\subsection*{Repeated feedback and pooling}

A separate experiment using Qwen2.5-0.5B tests two features of the proposed feedback process: accumulation and pooling.

To test accumulation, we hold the training questions and audience composition fixed at Dem90 or Rep90 while varying the number of synthetic users per item from 10 to 20 to 50. Increasing the number of users increases the number of repeated preference signals without introducing new political questions.

To test pooling, we fix the number of synthetic users at 20 per item and progressively replace group-specific feedback with feedback sampled from the overall OpinionQA population. Pooling levels are 0\%, 25\%, 50\%, 75\%, and 100\%. At 0\% pooling, Dem90 and Rep90 receive fully separate audience-specific feedback. At 100\% pooling, both adapters receive feedback from the same overall population distribution.

The Qwen2.5-0.5B accumulation and pooling experiments use one run per condition. They use 500 training questions, eight generated candidates per question, one DPO epoch, learning rate $5\times10^{-6}$, LoRA rank 16, LoRA alpha 32, dropout 0.05, target modules \texttt{q\_proj}, \texttt{k\_proj}, \texttt{v\_proj}, \texttt{o\_proj}, \texttt{gate\_proj}, \texttt{up\_proj}, and \texttt{down\_proj}, DPO $\beta=0.1$, batch size 1, gradient accumulation 8, maximum sequence length 1,024, bf16 precision, gradient checkpointing, and maximum gradient norm 1.0. The accumulation sweep uses seed 132 and 10, 20, or 50 synthetic users per item. The pooling sweep uses seed 133 and 20 synthetic users per item. A pooling level of 25\%, for example, means that 25\% of synthetic users are drawn from the overall OpinionQA distribution and the remaining 75\% from the model-specific 90/10 mixture. The pooled component is sampled independently for the two adapters; at 100\% pooling both adapters draw from the same overall distribution but do not share identical random draws.

\subsection*{Exploratory issue-domain analysis}

For exploratory analyses reported in the SI, we classify held-out questions using a transparent keyword taxonomy applied to the OpinionQA variable name, question text, and answer options. The domains are economic policy and welfare; culture, identity, and immigration; guns, crime, and public safety; climate and environment; health and science; government, democracy, and media; foreign policy; technology and privacy; and other social or personal issues.

Domain-level estimates are descriptive. Because the categories vary substantially in size, we retain question counts and model-specific estimates rather than interpreting the pooled estimates as a population-level ranking of domain susceptibility. We additionally examine whether DPO-induced movement is larger for questions with greater pre-existing partisan disagreement by correlating the original OpinionQA partisan gap with directional movement using Spearman rank correlation. 

The complete keyword taxonomy and item-level assignments are included in the replication archive. Briefly, keywords are matched against the variable name, question text, and answer options after lower-casing and concatenation. Categories are assigned by the first match in the following priority order: immigration/culture/identity; guns/crime/public safety; climate/environment; foreign policy; health/science; economic policy/welfare; government/democracy/media; technology/privacy; other social or personal issues. Thus, questions matching multiple domains are assigned to the earliest matching category in this ordered taxonomy.

\subsection*{Exploratory personalization experiment}

An exploratory experiment, reported in the SI, examines whether heterogeneous feedback is expressed as a general shift in model behavior or as a conditional response to user identity.

For each model family, we train one adapter on a balanced mixture of Democratic- and Republican-aligned preferences. Unlike the principal sorted-feedback design, each preference pair is accompanied by a prompt explicitly identifying the user as a Democrat or Republican. The adapter can therefore learn a conditional response policy in which the same model responds differently depending on the stated political identity of the user.

The experiment uses the Qwen2.5-1.5B, Mistral-7B, and GPT-OSS-20B base models, 500 training questions, eight candidate responses per question, and 50 synthetic users per item. Adapters train for two epochs with a learning rate of $2\times10^{-6}$. We evaluate neutral, Democratic-cued, and Republican-cued versions of 160 training and 160 disjoint test questions under the original and alternate prompt formats.

Cue-specific personalization is measured by (i) the proportion of questions on which Democratic and Republican identity cues produce different selected options and (ii) the difference in selected-option political margins between those cues. Each personalized adapter is compared with its untuned base model.

Because the experiment contains one training run per model, it is treated as exploratory evidence about possible modes of adaptation rather than as evidence for stable differences among model families.

\subsection*{Agent-based model}

The ABM represents the centrifugal alignment spiral in its minimal form. Politically heterogeneous users choose among initially similar AI models. Their choices determine the audiences from which each model receives feedback, and the models then update toward the feedback supplied by their own audiences. This process repeats over time.

\subsubsection*{Users and models}

User $i$ has a fixed political position

\begin{equation}
x_i \in [-1,1],
\end{equation}

where negative values denote left or liberal positions and positive values denote right or conservative positions. The benchmark population is drawn from a mixture of two normal distributions centered at $-0.4$ and $+0.4$, each with standard deviation 0.25, and clipped to $[-1,1]$.

The two mixture components have equal weight, so the benchmark population contains equal-sized left- and right-leaning groups in expectation.

Model $m$ has political position $z_m(t)\in[-1,1]$ at time $t$. Models begin with small random differences:

\begin{equation}
z_m(0)\sim
\mathcal{N}
\left(
0,\sigma_{\mathrm{initial}}
\right),
\end{equation}

with $\sigma_{\mathrm{initial}}=0.075$ in the benchmark simulation.

Initial model positions are clipped to $[-1,1]$, although clipping is rarely active under the benchmark initialization.

\subsubsection*{Model choice}

At each time step, every user chooses one model. The utility of model $m$ for user $i$ is

\begin{equation}
U_{im}(t)=
-\alpha
\left|
x_i-z_m(t)
\right|
+
\rho
\mathbf{1}
\left\{
c_i(t-1)=m
\right\}.
\end{equation}

where $\alpha$ controls sensitivity to political fit and $\rho$ is a persistence bonus for returning to the previously selected model. The persistence term captures habit, switching costs, lock-in, brand attachment, and other forces that can stabilize model choice.

Choice probabilities follow a multinomial logit:

\begin{equation}
P_{im}(t)=
\frac{
\exp[U_{im}(t)]
}{
\sum_k
\exp[U_{ik}(t)]
}.
\end{equation}

At the first choice step, users have no previous model, so the persistence term is omitted. The state recorded at $t=0$ contains only the initialized model positions; choices and feedback begin at $t=1$.

\subsubsection*{Audience feedback and model adaptation}

After choosing a model, each user generates a feedback signal

\begin{equation}
f_i(t)=x_i+\epsilon_i(t),
\qquad
\epsilon_i(t)\sim\mathcal{N}(0,\sigma_{\mathrm{feedback}}).
\end{equation}

Let $\bar{f}_m(t)$ denote the mean feedback from users who selected
model $m$. Models update according to

\begin{equation}
z_m(t+1)
=
z_m(t)
+
\eta\left[\bar{f}_m(t)-z_m(t)\right]
-
\lambda z_m(t),
\label{eq:update}
\end{equation}

where $\eta$ controls responsiveness to audience feedback and $\lambda$
regularizes models toward a shared neutral baseline. The update rule is
not intended as a literal simulation of production RLHF, DPO, or any
particular alignment pipeline. It abstracts the directional feature tested
experimentally in this study: politically structured feedback can move
model behavior toward the preferences represented in that feedback.
Implementation details and equilibrium properties are reported in the
SI Appendix.

\subsection*{Mapping LLM adaptation onto the ABM}

We use the fine-tuning experiments to define a plausible range for the model-responsiveness parameter $\eta$. The ABM time step is an abstract feedback-update cycle, not a single user interaction, calendar period, or literal production-model release. Under the benchmark ABM coding, the two user groups are centered at $-0.4$ and $+0.4$. A 90/10 audience therefore has an expected mean political position of magnitude

\begin{equation}
(2p-1)\mu
=
0.8\times0.4
=
0.32.
\end{equation}

The expected one-step gap between models adapting to opposing 90/10 audiences is therefore approximately

\begin{equation}
2\eta(0.32).
\end{equation}

Let $\bar{m}_{\mathrm{Dem90}}$ and $\bar{m}_{\mathrm{Rep90}}$ denote the average selected-answer political margins of the paired fine-tuned models on disjoint test questions. We define the effective update rate as

\begin{equation}
\hat{\eta}
=
\frac{
\bar{m}_{\mathrm{Dem90}}
-
\bar{m}_{\mathrm{Rep90}}
}{
2\times0.32
}.
\end{equation}

We use the disjoint test set because movement on these questions captures behavioral change beyond the examples used for alignment. The resulting effective update rates are approximately 0.017--0.019 under conservative alignment and 0.038--0.070 under stronger alignment. The tabulated anchoring values use a matched fixed-seed comparison; the principal headline results instead report five-seed means.

This mapping is an empirical bridge between the fine-tuning experiments and the ABM, not an estimate of a universal learning constant or a temporal update rate. It places the abstract model-update parameter on the same directional scale as the experimentally observed political movement. The simulations therefore use DPO results to anchor plausible responsiveness scenarios, while the repeated dynamics remain a model-based counterfactual.

\subsection*{Simulation design and outcome measures}

The benchmark ABM contains 5,000 users, four models, and 100 time steps. Unless otherwise noted, feedback noise is $\sigma_{\mathrm{feedback}}=0.20$, initial model dispersion is $\sigma_{\mathrm{initial}}=0.075$, feedback responsiveness is $\eta=0.05$, regularization is $\lambda=0.01$, choice sensitivity is $\alpha=4.0$, and persistence is $\rho=0$.

The main standard-condition simulations use 100 independent replications per condition. Mechanism checks use 20 replications per condition. Robustness checks use 20 replications per condition. DPO-anchored ABM scenarios use 50 replications per condition, and the full parameter sweep uses 100 replications per grid point. Replication $r$ uses random seed $12345+r$, so the same run index produces matched initial user populations and model positions across conditions unless a condition changes population or model-count parameters.

We track four outcomes. Model differentiation is the average pairwise distance among model positions:

\begin{equation}
\mathrm{Pol}(t)=
\frac{2}
{M(M-1)}
\sum_{m<k}
\left|
z_m(t)-z_k(t)
\right|.
\end{equation}

User segregation is the standard deviation across models of the mean political position of each model's current audience. Feedback skew is the standard deviation across models of mean audience feedback. Amplification is the ratio of final to initial model differentiation:

\begin{equation}
\mathrm{Amplification}
=
\frac{\mathrm{Pol}(T)}
{\mathrm{Pol}(0)}.
\end{equation}

Values above one indicate that initial differences among models have increased.

\subsection*{Mechanism checks and interventions}

We use three interventions and one persistence scenario to isolate the proposed mechanism.

In the \emph{shuffled-feedback} placebo, users choose models as usual, but their feedback signals are randomly permuted before feedback is aggregated by model. This preserves the distribution of user choices and feedback while breaking the link between the political composition of a model's audience and the feedback used to update that model.

In the \emph{pooled-feedback} intervention, all models receive the same feedback target: the mean feedback across the full user population at that time step. This removes model-specific differences in alignment signals while leaving user choice unchanged.

In the \emph{strong-regularization} intervention, $\lambda$ is increased from 0.01 to 0.05. This increases the pull toward the shared baseline.

In the persistence scenario, feedback remains audience-specific but $\rho$ is set to 1.99, the value implied by the pooled 71\% repeat-choice rate in the companion experiment.

The shuffled-feedback placebo is implemented by permuting individual feedback values across users after model choices are made and before model-level means are computed. This preserves the distribution of current audience memberships and the marginal distribution of feedback values, but breaks the association between a model's audience composition and the feedback signal used to update that model.

\subsection*{Parameter sweeps and transition analysis}

We vary user sensitivity to political fit ($\alpha$), model responsiveness to feedback ($\eta$), and regularization ($\lambda$) to identify the conditions under which initial model differences grow. Persistence in model choice ($\rho$) is examined in targeted mechanism and anchoring scenarios rather than in the full grid. The empirically anchored choice scenarios and fine-tuning-derived range of $\eta$ are overlaid on this parameter space.

The full grid is $\alpha\in\{0,0.25,0.5,1,1.5,2,3,4,5\}$, $\eta\in\{0,0.01,0.02,0.05,0.075,0.10,0.15\}$, and $\lambda\in\{0,0.01,0.05\}$, with $\rho=0$ throughout the grid. Each parameter combination is simulated for 100 independent replications. Boundaries are defined descriptively as the smallest $\alpha$ at which mean amplification across replications reaches or exceeds a specified value. The main boundary is amplification $>1$, indicating growth in model differentiation; supplementary boundaries use mean amplification of 2 and 5. This is a finite-horizon empirical boundary in a coarse parameter sweep, not a formal bifurcation threshold. The criterion is based on the mean, not on a confidence interval excluding the threshold. The benchmark $\alpha=4.0$ is near the upper end of the anchored range: it corresponds to the 19.9-point incentivized choice gap under the smaller assumed perceived distance ($d=0.05$), whereas $\alpha=2.0$ provides a less extreme anchored scenario (SI Appendix, Fig. S8).

\endgroup

\dataavail{Code and aggregate data supporting the analyses are available at \\ https://github.com/cssmodels/centrifugalalignmentspiral. The repository includes the agent-based model, DPO analysis code, survey-derived input data, summary tables, and figures. Model weights, fine-tuned adapters, raw model generations, and cluster-specific files are not redistributed but can be regenerated using the documented open-source models and dependencies. Survey-derived materials remain subject to the terms of their original sources.}
%

\bibliographystyle{unsrtnat}
\bibliography{pnas-sample}

\end{document}